\documentclass[sn-aps,Numbered]{sn-jnl} 
\usepackage{graphicx}
\usepackage{hyperref}
\usepackage{multirow}%
\usepackage[T1]{fontenc}
\usepackage[utf8]{inputenc}
\usepackage{lmodern}
\usepackage{booktabs}
\usepackage{animate}
\usepackage{bm}
\usepackage{subcaption}
\usepackage{amsmath,amssymb,amsfonts}%
\usepackage{amsthm}%
\usepackage{mathrsfs}%
\usepackage[title]{appendix}%
\usepackage{xcolor}%
\usepackage{textcomp}%
\usepackage{manyfoot}%
\usepackage{booktabs}%
\usepackage{algorithm}%
\usepackage{algorithmicx}%
\usepackage{algpseudocode}%
\usepackage{listings}%
\usepackage{mathtools}

\DeclarePairedDelimiterX\braket[2]{\langle}{\rangle}{#1 \delimsize\vert #2}

\usepackage[justification=centering]{caption}

\newcommand{\eexp}[1]{\text{e}^{#1}}
\newcommand{\di}[0]{\,\textrm{d}}

\newcommand{\cplxi}[0]{\text{i}}

\newcommand{\with}[0]{\quad\text{with}\quad}

\hypersetup{
    colorlinks=true,
    linkcolor=blue,
    filecolor=magenta,      
    urlcolor=cyan,
}

\begin{document}

\title[Article Title]{Spectral solver for Cauchy problems in polar coordinates using discrete Hankel transforms}

\author*[1,3]{\fnm{Rundong} \sur{Zhou} }\email{rundongz@student.chalmers.se}

\author[2]{\fnm{Nicolas} \sur{Grisouard}}\email{nicolas.grisouard@utoronto.ca}

\affil*[1]{\orgdiv{Division of Engineering Science}, \orgname{University of Toronto}, \orgaddress{\street{40 St.\ George Street}, \city{Toronto}, \postcode{M5S~2E4}, \state{Ontario}, \country{Canada}}}

\affil[2]{\orgdiv{Department of Physics}, \orgname{University of Toronto}, \orgaddress{\street{60 St.\ George Street}, \city{Toronto}, \postcode{M5S~1A7}, \state{Ontario}, \country{Canada}}}

\affil[3]{Now at\ \orgdiv{Department of Physics}, \orgname{Chalmers University of Technology}, \orgaddress{ \street{Kemig\aa rden 1}, \city{G\"oteborg}, \postcode{SE-412 96},  \country{Sweden}}}

\abstract{We introduce a Fourier-Bessel-based spectral solver for Cauchy problems featuring Laplacians in polar coordinates under homogeneous Dirichlet boundary conditions. We use FFTs in the azimuthal direction to isolate angular modes, then perform discrete Hankel transform (DHT) on each mode along the radial direction to obtain spectral coefficients. The two transforms are connected via numerical and cardinal interpolations. We analyze the boundary-dependent error bound of DHT; the worst case is $\sim N^{-3/2}$, which governs the method, and the best $\sim e^{-N}$, which then the numerical interpolation governs. The complexity is $O[N^3]$. Taking advantage of Bessel functions being the eigenfunctions of the Laplacian operator, we solve linear equations for all times. For non-linear equations, we use a time-splitting method to integrate the solutions. We show examples and validate the method on the two-dimensional wave equation, which is linear, and on two non-linear problems: a time-dependent Poiseuille flow and the flow of a Bose-Einstein condensate on a disk.}


\maketitle

\section{Introduction}
The Laplacian operator is associated with many important physical problems, including diffusion, the Schr\"odinger equation or the wave equation. It is more complicated under polar and spherical coordinates than in Cartesian coordinates. The literature has explored the use of the finite difference method for solving Poisson-type equations in cylindrical and spherical geometries~\cite{Lai2002, Lai2002_2}, and spectral approaches are often used to avoid coordinate singularities~\cite{Mohseni2000, Prochnow2017}. In this paper, we introduce a novel spectral solver suited for time-dependent problems featuring Laplacian in polar coordinates.

We consider a function $\psi$ on a unit disk $r \in [0,1]$ that satisfies the homogeneous Dirichlet boundary condition $\psi(1,\theta)=0$. The periodicity in the azimuthal direction allows for a decomposition of $\psi$ as a Fourier series, namely, 
\begin{equation} \label{eq:azimuthal decompose}
    \psi (r,\theta) = \sum_{q=-\infty}^{\infty} f_q(r) \eexp{\cplxi q \theta}, \with f_q(r) = \frac{1}{2\pi} \int_0^{2\pi} \psi (r,\theta) \eexp{-\cplxi q \theta} \di\theta
\end{equation}
and where $q$ denotes the angular mode number.
There are various choices of basis functions to decompose the radial function $f_q(r)$. Boyd and Yu~\cite{Boyd2011} list five spectral methods that use decompositions of the radial function. Among the listed basis functions, four are families of orthogonal polynomials, such as Zernike~\cite{Bhatia1954}, Logan-Shepp~\cite{Atkinson2010}, modified Chebyshev~\cite{Shen2000, Matsushima1995, Mohseni2000} or modified Jacobi polynomials~\cite{Vasil2015}. Most of them focus on solving boundary value problems and time-independent partial differential equations (PDEs) in polar coordinates, while Cauchy problems involving time evolution receive comparatively little attention.

To build a Laplacian solver suitable for time-dependent problems, instead of choosing orthogonal polynomial bases, we decided to use Bessel functions of the first kind as our basis to decompose the radial function $f_q(r)$, given by the Fourier-Bessel series~\cite{watson1995treatise},
\begin{equation}\label{eq:basisfunction}
    f_q(r) =  \sum_{j=1}^{\infty}a_{q,j} J_q\left(k_{q,j} r\right), \with a_{q,j} = \frac{2}{J_{q+1}^2(k_{q,j})} \int_0^1 r f_q(r) J_q(k_{q,j}r)\di r.
\end{equation}
$J_q$ is the $q^{\text{th}}$-order Bessel function of the first kind and $k_{q,j}$ denotes its $j^{\text{th}}$ non-negative zero. Temme~\cite{Temme1979} provides a fast and accurate algorithm to compute these roots.

The Fourier-Bessel basis (i.e., the combination of the azimuthal Fourier basis of Eqn.~\ref{eq:azimuthal decompose} and radial Bessel basis of Eqn.~\ref{eq:basisfunction}) has an algebraic rate of convergence $\lesssim  1/N^{5/2}$ for functions satisfying homogeneous Dirichlet boundary conditions~\cite{Boyd2011}, where $N$ is the number of basis functions used for approximation.
This compares unfavourably with many orthogonal polynomial bases which have an exponential rate.
However, Fourier-Bessel modes are eigenfunctions of the Laplacian operator, and the boundary conditions are enforced by the basis functions themselves. Such virtues ensure that a pure spectral scheme can be applied at each iteration. It makes the Fourier-Bessel basis a competitive choice for solving time-dependent initial value problems associated with Laplacians under homogeneous Dirichlet conditions in polar coordinates. We further notice a recent related work~\cite{marshall2022fast} on obtaining Fourier-Bessel coefficients but considering time-independent problems on a Cartesian sampling grid.

To decompose the radial function $f_q(r)$ into Fourier-Bessel series, we use the discrete Hankel transform (DHT). It was first introduced mathematically by Johnson~\cite{Johnson1987}, then independently re-invented by Yu et al.~\cite{Yu:98} for the zeroth order mode ($q=0$) and Guizar-Sicairos et al.~\cite{2004_JOSAA_Guizar-SicairosG} for all integer orders. Finally, It was categorized by Baddour~\cite{Baddour2019DHT} as a discrete variation of general Fourier transform.

This paper is organized as follows: we first reformulate the DHTs as pseudospectral collocation method by introducing their discrete inner products, quadrature weights, pseudospectral grid points, and cardinal interpolations in Section~\ref{sec:DHT} and~\ref{sec:interpolation}. We analyze error and complexity and discuss the boundary dependency of the error bound in Section~\ref{sec:error_and_runtime}. Then we introduce a systematic way to apply the method to compute Laplacian and solve nonlinear time-dependent equations in Section~\ref{sec:laplacian}. In Section~\ref{sec:examples} we show three examples: the linear 2-D wave equation where we test the convergence of the method numerically (\S~\ref{sec:membrane}), a time-dependent Poiseuille flow equation (\S~\ref{sec:Poiseuille}), and the Gross-Pitaevskii equation (\S~\ref{sec:BEC}). In Section~\ref{sec:conclusion} we conclude the paper and discuss potential improvements.

\section{Method}

\subsection{Discrete Hankel transform and pseudospectral grid points} \label{sec:DHT}

Our first step is to reformulate the discrete Hankel transforms as pseudospectral collocation methods. The optimal pseudospectral grid points are the zeros of the $(N+1)^\text{th}$ basis function $\phi_{N+1}(r)$ \cite[\S\,4.3]{boyd2013chebyshev}, where $N$ is the total number of basis functions, or radial modes, used to approximate the radial function $f_q(r)$. The basis functions of a $q^\text{th}$-order Hankel transform are then
\begin{equation}
    \phi_{q,i}(r) = J_q\left(k_{q,i} r\right).
\end{equation}
Thus, the $(N+1)^\text{th}$ basis function is $J_q(k_{q,N+1} r)$, and we choose its pseudospectral grid points to be
\begin{equation}\label{eq:grid}
    \{r_{q,i}\}_{i=1,2,...,N} = \frac{k_{q,i}}{k_{q,N+1}}.
\end{equation}
We can now define the discrete inner product as \cite[\S\,3.1.4]{Shen2011}
\begin{equation}
    (f,\: g)_q \equiv \sum_{i=1}^N w_{q,i} f(r_{q,i})g(r_{q,i}),\quad\text{where}\quad w_{q,i} = \frac{2}{k_{q,N+1}^2 J_{q+1}^2(k_{q,i}) }
\end{equation}
are the quadrature weights.

Bessel functions are orthogonal under our discrete inner product, namely \cite{Johnson1987}
\begin{equation}\label{eq:orthogonality}
    \frac{2}{J_{q+1}^2(k_{q,m})}\Big(J_q(k_{q,m}r),\: J_q(k_{q,n}r)\Big)_q = \delta_{mn}.
\end{equation}
The discrete Hankel transform of the radial function $f_q(r)$ is now defined through the discrete inner product as
\begin{equation}
    F_{q,j} = \Big(f_q(r), J_q(k_{q,j}r)\Big)_q.
\end{equation}
By substituting the definition of discrete inner product and pseudospectral grid points $\{r_{q,i}\}$, the forward discrete Hankel transform formula is then given by
\begin{equation}\label{eq:forward}
    F_{q,j} = \frac{2}{k_{q,N+1}^2} \sum_{i=1}^{N}\frac{J_q\left(k_{q,i} k_{q,j} / k_{q,N+1}\right)}{J_{q+1}^2\left(k_{q,i}\right)}f_q\left( r_{q,i}\right).
\end{equation}
The value of the radial function $f_q(r)$ at each pseudospectral grid point $r_{q,i}$ can be recovered by the Fourier-Bessel series Eqn.~\eqref{eq:basisfunction}, namely,
\begin{equation}\label{eq:backward}
    f_q\left( r_{q,i}\right) = 2\sum_{j=1}^{N} \frac{J_q\left(k_{q,i} k_{q,j} / k_{q,N+1}\right)}{J_{q+1}^2\left(k_{q,j}\right)}F_{q,j}.
\end{equation}
The transform formulae Eqns.~\eqref{eq:forward} and \eqref{eq:backward} demonstrate symmetry and can be implemented as matrix multiplications. Introducing
\begin{equation}
    f_{q,i} \equiv f_q\left( r_{q,i}\right) \quad\text{and} \quad M_{q,ij} \equiv \phi_{q,i}(r_{q,j})w_{q,i} = \frac{2 J_q\left(k_{q,i} k_{q,j} / k_{q,N+1}\right)}{k_{q,N+1}^2 J_{q+1}^2\left(k_{q,i}\right)},
\end{equation}
we obtain
\begin{equation}
    \vec{F}_q = \textbf{M}_q \cdot\vec{f}_q \: ; \quad \vec{f}_q = k_{N+1}^2 \textbf{M}_q \cdot\vec{F}_q,
\end{equation}
where $\textbf{M}_q$ is unitary, i.e., $\textbf{M}_q \textbf{M}_q^{-1} = \textbf{I}$.
The discrete Hankel transform can be categorized as a Matrix Multiplication Transformation \cite[\S\,10.4]{boyd2013chebyshev}, and it demonstrates the spectral methods' virtue of one-to-one correspondence between spectral coefficients $F_{q,j}$ and pseudospectral grid points $\{f_q(r_{q,i})\}$.  

\subsection{Interpolation and cardinal functions} \label{sec:interpolation}
\begin{figure}
    \centering
    \begin{subfigure}[b]{0.425\textwidth}
         \centering
         \includegraphics[width=\textwidth]{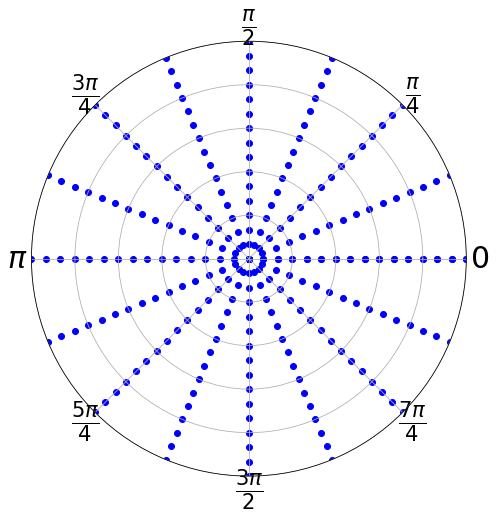}
         \caption{FFT grid}
         \label{fig:FFT_grid}
     \end{subfigure}
     \hfill
     \begin{subfigure}[b]{0.47\textwidth}
         \centering
         \includegraphics[width=\textwidth]{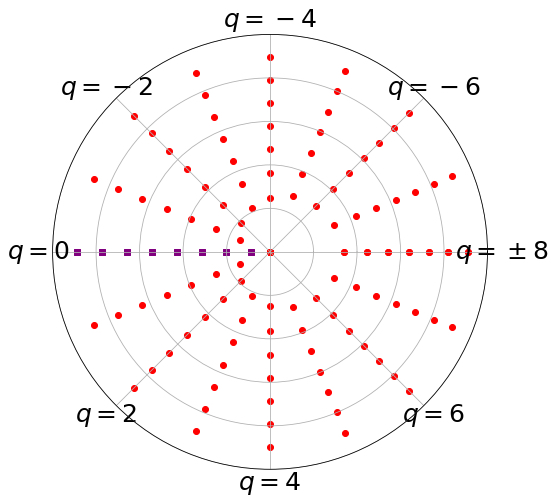}
         \caption{DHT grid}
         \label{fig:DHT_grid}
     \end{subfigure}
    \caption{Difference between the evenly sampled FFT grid on $\psi(r,\theta)$ and the pseudospectral $\{r_{q,i}\}$ grid for the DHTs in the case of $N=8$. In panel (b), the sampling points for $q=0$ (marked as purple squares) do not contain the origin since $r_{0,1}>0$.}
    \label{fig:grid}
\end{figure}

To obtain the spectral coefficients $a_{q,j}$ of Eqn.~\eqref{eq:basisfunction}, we perform a Fast Fourier Transform (FFT) along the azimuthal direction followed in the radial direction by one DHT for each angular mode $q$. The pseudospectral grid points $\{r_{q,i}\}$ in Eqn.~\eqref{eq:grid} defined for the DHTs are not evenly spaced and vary with the angular mode $q$.  
Instead of using Baddour and Yao's unevenly sampled grid~\cite{Baddour2019, Yao2020}, which causes unwanted artifacts in the center of the domain, we choose an evenly sampled grid that is finer in the radial direction before performing the azimuthal FFT (hereafter referred to as the FFT grid), as visualized in Figure~\ref{fig:FFT_grid}. We then numerically interpolate each radial function $f_q$ on the corresponding DHT grid $\{r_{q,i}\}_{i=1,\dots, N}$ to obtain $\{f_q(r_{q, i})\}_{i=1,\dots, N}$, as shown in Figure~\ref{fig:DHT_grid}. 

To achieve optimal numerical accuracy for a given $N$, the number of radial sampling points of the finer FFT grid should be at least $2N$ (see \S\,\ref{sec:error}) The number of azimuthal sampling points is twice that of the desired angular modes $q$. 
Conversely, when we transform from the spectral domain $a_{q,j}$ back to the physical domain $\psi(r,\theta)$, we perform inverse DHTs followed by an inverse FFT. 
We use the cardinal functions of Fourier-Bessel series to interpolate from the pseudospectral grid $\{r_{q,i}\}$ to the finer FFT grid~\cite{Baddour2019DHT}, namely,
\begin{equation} \label{eq:cardinal}
    f_q(r) = \sum_{i=1}^\infty f_q(r_{q,i})C_{q,i}(r), \quad \textrm{where}\quad C_{q,i}(r) = \frac{2k_{q,i}}{k_{q,i}^2 - r^2k_{q,N+1}^2} \frac{J_q(rk_{q,N+1})}{J_{q+1}(k_{q,i})}.
\end{equation}
We discuss the error introduced by these two interpolations in the following section.

\subsection{Error analysis and complexity} \label{sec:error_and_runtime}
\subsubsection{Numerical interpolation} \label{sec:error}
The numerical interpolation connecting the finer FFT grid to the pseudospectral grid $\{r_{q,i}\}$ introduces an error that largely depends on the resolution of the finer FFT grid and the choice of numerical method. We choose the cubic spline interpolation, whose global error $E_I$ to approximate the radial function $f_q(r)$ is bounded by \cite{HALL1976105}
\begin{equation*}
    E_I \leq \frac{5}{384} \left|f_q^{(4)}\right| h^4,
\end{equation*}
where $h$ is the size of the radial interval of the FFT grid. According to our sampling strategy explained in \S\,\ref{sec:interpolation}, there are $2N$ radial FFT sampling points and hence $h=(2N)^{-1}$. By assuming that the radial function is well-behaved, the numerical interpolation error goes as
\begin{equation} \label{eq:numerical}
    E_I(N) \sim N^{-4}.
\end{equation}

\subsubsection{Cardinal interpolation, general error of DHT} \label{sec:cardinal_error}
This section discusses the error introduced by the cardinal interpolation Eqn.~\eqref{eq:cardinal}, which is also the general error associated with using discrete Hankel transforms. According to the convergence theorem of Fourier-Bessel series, an arbitrary radial function $f_q(r)$ can be fully recovered by Eq.~\eqref{eq:basisfunction}. We denote its truncation up to the $N^\text{th}$ term by
\begin{equation}
    f_q^N (r) \equiv \sum_{j=1}^{N}a_{q,j} J_q\left(k_{q,j} r\right).
\end{equation}
The truncation error $E_T$ is then
\begin{equation}
    E_T (q,N) \equiv \left|f_q^N (r) - f_q (r)\right| \leq \sum_{j=N+1}^{\infty}|a_{q,j}| \sup_{x>0} \left\{|J_q\left(x\right)|\right\}.
\end{equation}

The $q^{\textrm{th}}$-order Bessel function of the first kind is characterized by~\cite{Landau2000}
\begin{equation*}
    |J_q(x)| < b_q q^{-1/3},
\end{equation*}
where the $b_q$'s are such that $b_1 \simeq 0.58$, increasing monotonically to around 0.67 as $q$ tends to infinity. We can conclude that, for orders $q \geq 1$, the truncation error is 
\begin{equation}
    E_T (q,N) < b_q q^{-1/3}\sum_{j=N+1}^{\infty}|a_{q,j}|.
\end{equation}
For $q=0$, $\sup_{x} \left\{|J_0\left(x\right)|\right\} = 1$ and therefore,
\begin{equation}
    E_T (0,N) \leq \sum_{j=N+1}^{\infty}|a_{0,j}|.
\end{equation}
For the algebraically-converging Fourier-Bessel series, the magnitude of the truncation error is given by~\cite[\S\,2.13]{boyd2013chebyshev}
\begin{equation}\label{eq:error}
     E_T (q=0,N) \sim  N|a_{0,N}|, \quad \textrm{and} \quad E_T (q\geq 1,N)\sim  b_q q^{-1/3}N|a_{q,N}|.
\end{equation}

Now we denote the interpolated function approximated by $N$ radial modes via DHT's cardinal functions, i.e., Eqn.~\eqref{eq:cardinal}, as $S_q^N(r)$. It agrees with the original radial function $f_q(r)$ on the pseudospectral grid points $\{r_{q,i}\}$ defined in Eqn.~\eqref{eq:grid}, namely,
\begin{multline}
    S_q^N(r) = \sum_{i=1}^N f_q(r_{q,i})C_{q,i}(r) = 2\sum_{j=1}^N \frac{F_{q,j}}{J_{q+1}^2\left(k_{q,j}\right)} J_q\left(k_{q,j} r\right), \quad \\
    \textrm{and} \quad \left\{ S_q^N(r_{q,i})\right\}_{i=1,\dots, N} = \left\{f_q(r_{q,i})\right\}_{i=1,\dots, N} ,
\end{multline}
where the $F_{q,j}$'s are the result of the DHT, i.e., Eqn.~\eqref{eq:forward}. Note that because $J_q(k_{q,j})=0$ by definition of the $k_{q,j}$'s, the Dirichlet boundary condition is automatically enforced on the radial function, i.e., $f_q(1)=0$. This solves the problem that the pseudospectral grid does not sample the boundary $r=1\notin \{r_{q,i}\}$. Using the discrete orthogonality relation Eqn.~\eqref{eq:orthogonality}, we show in \ref{sec:proof} that the discrete Hankel transform on $N$ pseudospectral grid points produces the exact first $N$ Fourier-Bessel coefficients $a_{q,j}$. In other words,
\begin{equation} \label{eq:equality}
 F_{q,j}= \frac{J_{q+1}^2(k_{q,j})}{2}a_{q,j} =  \int_0^1 r f_q(r) J_q(k_{q,j}r)\di r  , \quad \textrm{and thus} \quad S_q^N(r) = f_q^N(r).
\end{equation}
Therefore, the discrete Hankel transform introduces the same amount of error as truncating an infinite Fourier-Bessel series would (i.e., $E_T$). From Eqn.~\eqref{eq:error} and the convergence rate of Fourier-Bessel series for functions satisfying the homogeneous Dirichlet condition, $a_{q,N} \lesssim N^{-5/2}$, we obtain the general error of DHTs,
\begin{equation} \label{eq:total_error}
    E_D (q=0,N) \lesssim  N^{-3/2}, \quad \textrm{and} \quad E_D (q\geq 1,N)\lesssim  b_q q^{-1/3} N^{-3/2}.
\end{equation}

Note that `$\lesssim$' above denotes the worst case, whereas the actual error may be smaller depending on the radial function $f_q(r)$. Indeed, the convergence rate of Fourier-Bessel series depends mainly on the boundary behavior of the function~\cite{Boyd2011}. The series has a faster convergence rate and hence a smaller error if the function satisfies the following condition at the boundary up to its $p^{\textrm{th}}$ derivative:
\begin{equation}
    f_q(1) = f_q'(1) = f_q''(1) = \dots = f_q^{(p)}(1) = 0. \label{eq:p}
\end{equation}
The higher $p$, the faster the convergence rate. Following Boyd's result, such an error bound is given by $E_D (p,N) \sim  N^{-2p-3/2}$. That is, for the worst-case scenario in which $p=0$, i.e., in which the homogeneous Dirichlet condition is satisfied for the function and not its first derivative, Eqn.~\eqref{eq:total_error} gives the error.  For the best case, $p=\infty$, we have an exponential rate of convergence. The DHT error is then governed by $E_D (N) \sim  \eexp{-N}$, although this occurs very rarely.

\subsubsection{Interpolation vs.\ DHT errors}

We saw that there were two competing sources of error in our method, the interpolation error $E_I$ (Eqn.~\ref{eq:numerical}) and the DHT error $E_D$ due to the truncation of the infinite Fourier-Bessel series (Eqn.~\ref{eq:total_error}). Because $E_I \sim N^{-4}$ vs.\ $E_D \sim N^{-3/2}$ in the worst-case scenario described above (case $p=0$ in Eqn.~\ref{eq:p}), the truncation error dominates. However, if the function is such that $p$ is large (cf.\ Eq.\ \ref{eq:p}), numerical interpolation can become the main source of error.

All the error analyzes conducted so far are in the radial direction, associated with decomposing the radial function $f_q(r)$. Indeed, in the azimuthal direction, FFTs have an exponential rate of convergence; thus, their error is negligible compared to the largest of the interpolation or DHT errors.

\subsubsection{Runtime analysis} \label{sec:complexity}
Since the main purpose of the method is to solve time-dependent initial value problems, multiple transforms are executed for a large number of time iterations. Therefore, the complexity of the operations is of great concern. We now analyze how the number of operations scales with the spatial resolution. Consider the function $\psi(r,\theta)$, sampled with $2N$ azimuthal and radial points, where $N$ is the number of radial modes. FFTs on such grids have a complexity of $4N^2\log(2N)$.

After the Fourier transform, we perform a cubic spline interpolation (complexity $N$ \cite{cubic_spline_complex}) within each angular mode $q$, where the $q$'s are integers ranging from $-N$ to $N$. The complexity of the numerical interpolation with non-periodic boundary conditions is therefore $2N^2$.

The discrete Hankel transform is a multiplication of a $N$-vector with a $N \times N$ dense matrix. The total complexity is then $2N^3 $.
Thus, the DHT operation governs the forward transform from $\psi(r, \theta)$ to spectral coefficients $\{a_{q,j}\}$, the complexity is given by
\begin{equation*}
    T_{\textrm{forward}}(N) \sim 2N^3.
\end{equation*}

Conversely, for the backward transform, the FFT and DHT terms remain the same. We use the cardinal interpolation introduced in \S\,\ref{sec:interpolation}, instead of cubic splines. It is equivalent to a $2N \times N$ matrix with a $N$ vector multiplication. Thus, its complexity is $ 2N^3 $.
Hence, both cardinal interpolation and DHT terms govern the backward transform, the complexity is then
\begin{equation*}
    T_{\textrm{backward}}(N) \sim 4N^3.
\end{equation*}
Thus, the total complexity associated with the method is $O[N^3]$. The matrices for DHT and cardinal interpolation Eqn.~\eqref{eq:forward}, \eqref{eq:backward} and \eqref{eq:cardinal}, can be pre-computed outside temporal loops and do not affect the complexity when solving initial value problems.

\subsection{Applying the Laplacian operator and spectral time scheme} \label{sec:laplacian}
Bessel functions are the eigenfunctions of the Laplacian operator in polar coordinates. That is, for the function $\psi(r,\theta)$ decomposed under our method with $N$ azimuthal and radial modes, namely,
\begin{equation}
    \psi (r,\theta) = \sum_{q=-N}^{N} \sum_{j=1}^{N} a_{q,j} \eexp{\cplxi q \theta}J_q(k_{q,j} r),
\end{equation}
the Laplacian of this function becomes
\begin{equation}
    \nabla^2 \psi (r,\theta) = \sum_{q=-N}^{N} \sum_{j=1}^{N} -k_{q,j}^2 a_{q,j} \eexp{\cplxi q \theta}J_q(k_{q,j} r).
\end{equation}
The set of basis functions remains unchanged. A pure spectral time scheme can be applied here and the propagators are the Bessel roots $k_{q,j}$. The implementation is straightforward: we use the transform method introduced in previous sections to obtain the spectral coefficients $a_{q,j}$, then apply a time scheme (the operator splitting in our case), and finally transform back to the physical domain. The homogeneous Dirichlet boundary condition is enforced by the Bessel functions without extra operations. 

To solve PDEs associated with operators other than the Laplacian, we use the operator splitting method~\cite{operator_split}. It can be applied to differential equations of the form
\begin{equation}
    \frac{\partial \psi}{\partial t} = c \nabla^2 \psi + \mathfrak{D}[\psi],
\end{equation}
where $c$ is a constant, by creating an intermediate state $\psi_{\textrm{int}}$ that would solve $\partial_t \psi_{\textrm{int}} = c \nabla^2 \psi$. In our case,
\begin{equation} \label{eq:time_scheme}
    \psi_{\textrm{int}} = \sum_{q=-N}^{N} \sum_{j=1}^{N} \left\{ a_{q,j} \eexp{- c k_{q,j}^2 \Delta t} \right\}\eexp{\cplxi q \theta}J_q(k_{q,j} r)
\end{equation}
where $\Delta t$ is the time step. The other operator $\mathfrak{D}$ now acts on $\psi_{\textrm{int}}$, and can be treated with other numerical methods. Due to the nature of this spectral time scheme, calculating $\psi_{\textrm{int}}$ is mathematically exact, the sole source of error is from approximating the function via DHTs. For PDEs associated with second-order time derivatives, readers can refer to the first example illustrated in \S\,\ref{sec:membrane}.
The method is fast, accurate, easy to program, and it has many applications in polar coordinates associated with Laplacian operators,  as we demonstrate next.

\section{Examples} \label{sec:examples}

\subsection{Drum beats} \label{sec:membrane}
Vibrations of an elastic membrane are governed by the 2-D wave equation, that is,
\begin{equation}\label{eq:wave}
    \frac{\partial^2 u}{\partial t^2} = c^2 \nabla^2 u,
\end{equation}
where $u(r, \theta, t)$ denotes the transverse displacement of the membrane and $c$ is the wave speed. It has the following general solution in 2-D polar coordinates under homogeneous Dirichlet conditions,
\begin{equation} \label{eq:membrane_solution}
    u(r, \theta, t) = \sum_{q=-\infty}^{\infty} \sum_{j=1}^{\infty} \left[a_{q,j} \cos(c k_{q,j} t) + \frac{b_{q,j}}{c k_{q,j}} \sin( c k_{q,j} t)\right] \eexp{\cplxi q \theta} J_q(k_{q,j} r),
\end{equation}
where $a_{q,j}$ and $b_{q,j}$ are the Fourier-Bessel coefficients of the initial displacement and velocity fields, namely,
\begin{equation} \label{eq:membrane_coefficient}
    \begin{cases}
            \sum_{q} \sum_{j} a_{q,j} \eexp{\cplxi q \theta}J_q(k_{q,j} r) = u|_{t=0} \\

        \sum_{q} \sum_{j} b_{q,j} \eexp{\cplxi q \theta}J_q(k_{q,j} r) = \partial_t u|_{t=0}, 
    \end{cases}
\end{equation}
can be directly obtained by the method introduced in this paper. This example demonstrates the great advantage of spectral methods with eigenfunction bases to solve linear equations, because the state at any arbitrary future time $t$ can be determined directly from the initial conditions. The time scheme is fully spectral and no time evolution is required.

\begin{figure}
    \centering
    \includegraphics[width = 90 mm]{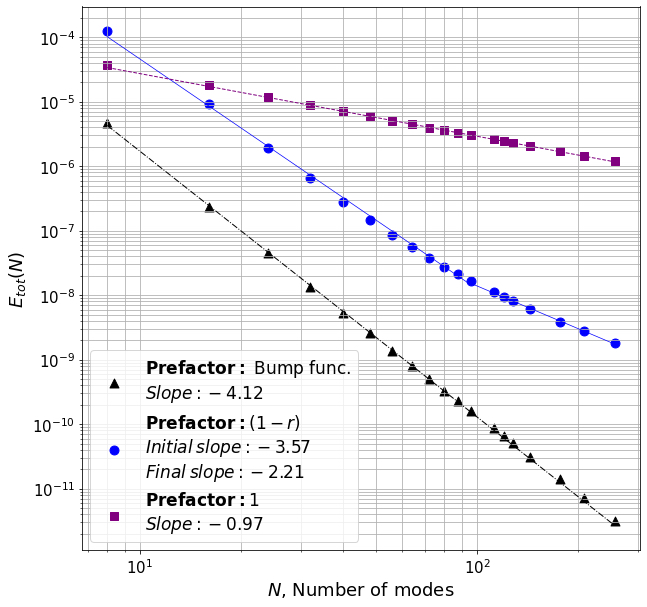}
    \caption{Comparison of the method's convergence rates on Eqn.~\eqref{eq:membrane_initial} with different prefactors.}
    \label{fig:membrane_error_fit}
\end{figure}

To validate the method, we construct a circular membrane with initial displacement consisting of two off-centered Gaussians with a prefactor term $\alpha (r)$ such that we can control its boundary behavior, and no initial velocity (in which case $b_{q,j} = 0$ $\forall (q, j)$). In other words,
\begin{equation}\label{eq:membrane_initial}
\begin{split}
        u|_{t=0} = \alpha(r) \Big(\eexp{ -15[r^2 + r_0^2 - 2rr_0\cos(\theta - \theta_0)]} &- \eexp{ -15[r^2 + r_1^2 - 2rr_1\cos(\theta - \theta_1)]}\Big)\\
        &\quad\text{and}\quad \partial_t u|_{t=0} = 0,
\end{split}
\end{equation}
where $r_0, \theta_0$, $r_1$, and $\theta_1$ locate the centers of the two Gaussian peaks, respectively,
\begin{equation*}
   \begin{cases} 
   r_0 = r_1 = 0.3 \\
   \theta_0 = \theta_1 -\pi = 1.
   \end{cases}
\end{equation*}
We perform pairs of forward and backward transforms without time propagation with different $N$'s (number of azimuthal and radial modes) on Eqn.~\eqref{eq:membrane_initial} to check the convergence of the method. We define the total error as
\begin{equation}
    E_{\textrm{tot}} = \int \frac{|u_{\textrm{transformed}}-u|}{\pi\max (u)}  \di\mathbf{a},
\end{equation}
where $\di\mathbf a = r\di r \di \theta$ is the infinitesimal patch of area.
Figure~\ref{fig:membrane_error_fit} shows $E_{\textrm{tot}}$ vs.\ $N$ (number of radial and azimuthal modes) with three different prefactors $\alpha(r)$.

For $\alpha(r) = 1-r$ (blue dots), the convergence rate starts at $\sim N^{-3.57}$, then decreases to $\sim N^{-2.21}$ when $N$ reaches $10^2$. The displacement field satisfies the Dirichlet conditions, but its radial derivative is non-zero at the boundary, and thus $p=0$, as discussed in \S\ref{sec:cardinal_error}. However, the derivative is small: $\partial_r u |_{r=1}\simeq 6.4\times 10^{-4}$ according to our calculation. This explains why the convergence rate here is faster than the theoretical bound; that is, the method is less sensitive to the function's higher-order derivatives' boundary conditions when the resolution is low ($N$ is small). It is not practical to check the convergence at very large $N$, but we can predict that it will eventually reach the worst-case error bound $E_D \sim N^{-3/2}$ as $N\rightarrow\infty$. This is a positive feature, since in practical usages we generally choose $N \sim 10^2$, and for such $N$'s, the method converges faster and thus has less error than the theory. 
 
For $\alpha(r) = \exp \left[-(1-x^2)^{-1}\right]$, namely, the bump function (black triangles), the convergence rate is given by $\sim N^{-4.12}$, and shows no sign of decreasing.
The bump function has the property that all its derivatives cancel at $r=1$, corresponding to $p =\infty$ in \S\,\ref{sec:cardinal_error} and the Fourier-Bessel series of the resulting initial condition has an exponential convergence rate. Thus, the error bound of the method is no longer governed by the DHT, but by the cubic spline interpolation $E_I \sim N^{-4}$, \S~\ref{sec:error}, which agrees with our numerical result.
 
Finally, we let $\alpha(r) = 1$, thus the displacement field no longer satisfies the homogeneous Dirichlet conditions. The convergence rate (purple squares) drops below the theoretical worst-case error bound to around $\sim N^{-1}$, as the theory predicts.

We have verified our statements on the error bound and demonstrated the boundary-dependent nature of the method. When applying the method to a function, we should always consider the boundary behavior to achieve the desired accuracy.

\begin{figure}
\centering
    \includegraphics[width = 1\textwidth]{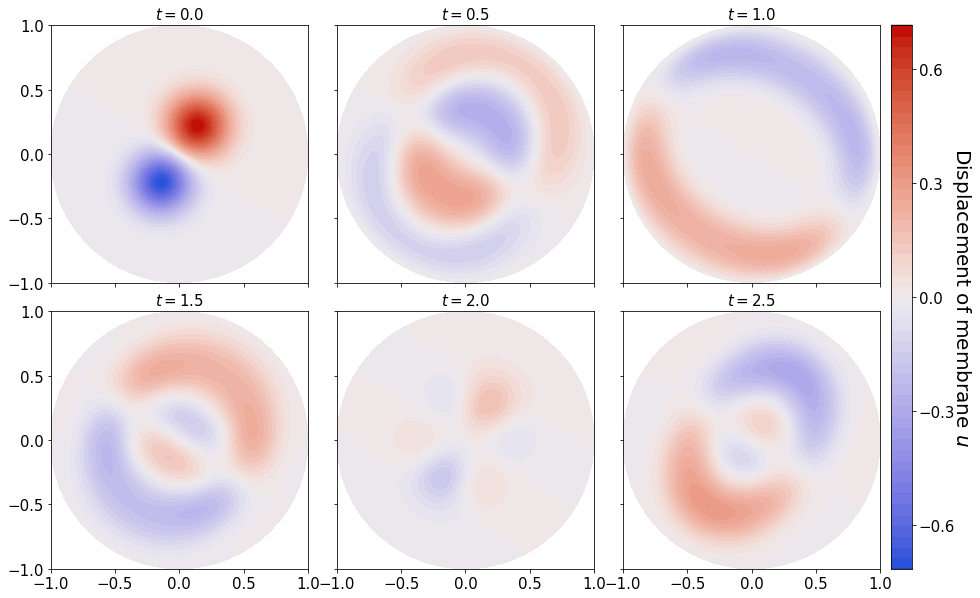}

    \caption{Solving the wave equation \eqref{eq:wave} in polar coordinates. Top-left panel: Initial condition of the membrane displacement $u(r, \theta, t=0)$, consisting of two off-centred Gaussians with the prefactor $\alpha = 1-r$ (cf.\ Eq.\ \ref{eq:membrane_initial}). Other panels: Membrane displacements at $t=\{0.5, 1.0, 1.5, 2.0, 2.5\}$.}
    \label{fig:Membrane}
\end{figure}

Figure \ref{fig:Membrane} shows an example using the method to solve the evolution of \eqref{eq:membrane_initial} with $\alpha(r)=1-r$ at six different times.
For simplicity, 
we set the sound speed to unity ($c=1$). A remarkable feature of this example is that the system has an infinite period if we have more than one mode, since the Bessel roots $k_{q,j}$'s are transcendental and have no multiple of one another~\cite{Siegel2014}.


\subsection{Impulse start of a Poiseuille pipe flow \label{sec:Poiseuille}}
Our first non-linear example is the time-dependent Poiseuille pipe flow described by
\begin{equation} \label{eq:poiseuille}
     \partial_t u = G(r,\theta,t) + Re^{-1} \nabla^2 u \quad\text{and}\quad u|_{r=1}=0,
\end{equation}
which solves for the velocity field $u$ of an incompressible, homogeneous laminar flow across a unit circle, where $G(r,\theta,t)$ is the normalized pressure gradient and $Re$ is a constant called the Reynolds number of the flow.
The initial condition is $u|_{t=0}=0$ everywhere on the circle. 

First, we consider the simplest case where the pressure gradient $G$ is constant throughout the domain at positive times. The problem now has radial symmetry and only the $q=0$ angular mode is involved, reducing to a one-dimensional (1D) problem, whose analytical solution is \cite[\S\,4.3]{Batchelor2000}
\begin{equation} \label{eq:poiseuille_analytical}
    u(r,t) = \left[\frac{1-r^2}{4} - 2 \sum_{i=1}^{\infty} \frac{J_0(k_{0,i} r)}{k_{0,i}^3 J_1(k_{0,i})} \exp\left(-\frac{k_{0,i}^2 t}{Re}\right)\right]GRe.
\end{equation}
Note that as $t \rightarrow \infty$, we recover the classical stationary Poiseuille flow $u_\infty = (1-r^2)GRe/4$.

\begin{figure}[htbp]
\centering
    \begin{subfigure}[b]{0.32\textwidth}
         \centering
         \includegraphics[width=\textwidth]{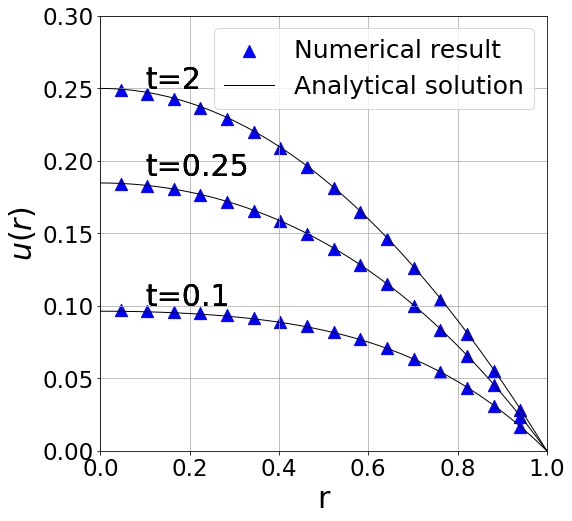}
         \caption{Time evolution.}
         \label{fig:Poiseuille-1D_evolution_16}
     \end{subfigure}
     \hfill
     \begin{subfigure}[b]{0.32\textwidth}
         \centering
         \includegraphics[width=\textwidth]{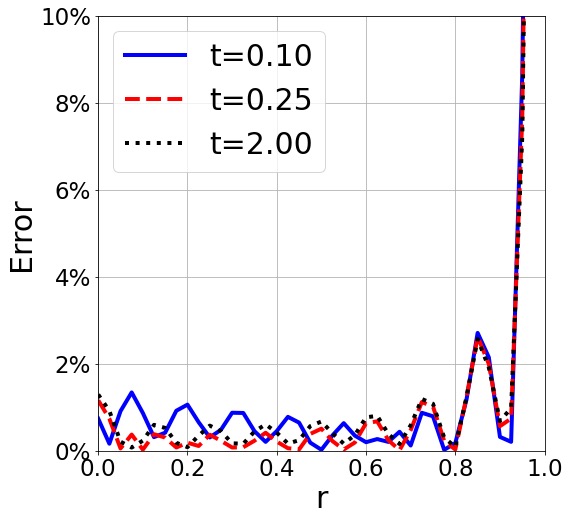}
         \caption{Relative percentage error, $\Delta t = 10^{-3}$.}
         \label{fig:Poiseuille-1D_error_16}
     \end{subfigure}
     \hfill
     \begin{subfigure}[b]{0.32\textwidth}
         \centering
         \includegraphics[width=\textwidth]{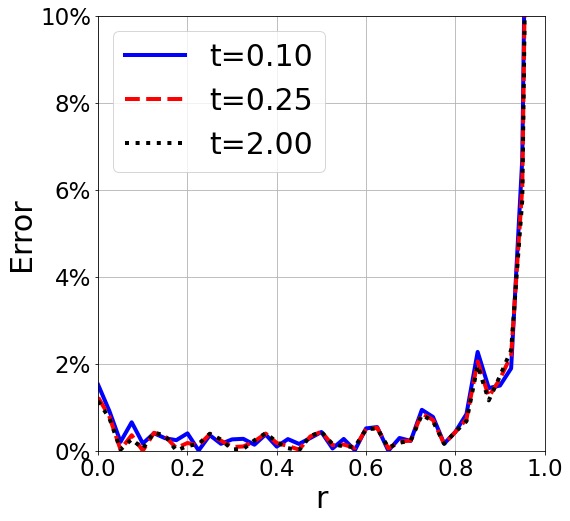}
         \caption{Relative percentage error, $\Delta t = 10^{-4}$.}
         \label{fig:Poiseuille-1D_error_16_dt=10^-4}
     \end{subfigure}

    \caption{Simulation of the impulse start of a radially-symmetric Poiseuille pipe flow with $N=16$ radial modes.}
    \label{fig:Poiseuille_1D_16}
\end{figure}

\begin{figure}[htbp]
\centering
    \begin{subfigure}[b]{0.32\textwidth}
         \centering
         \includegraphics[width=\textwidth]{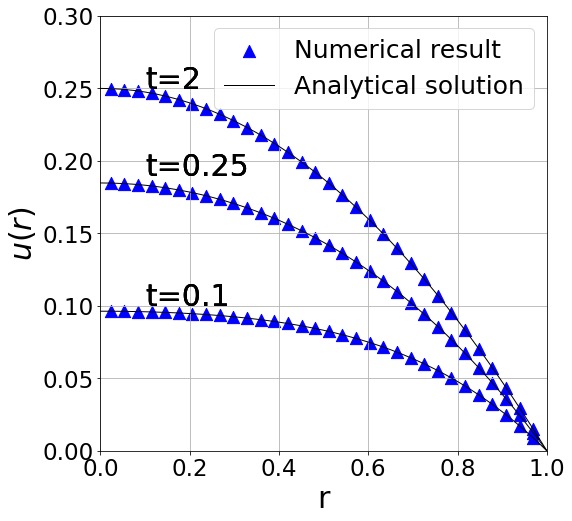}
         \caption{Time evolution.}
         \label{fig:Poiseuille-1D_evolution_32}
     \end{subfigure}
     \hfill
     \begin{subfigure}[b]{0.32\textwidth}
         \centering
         \includegraphics[width=\textwidth]{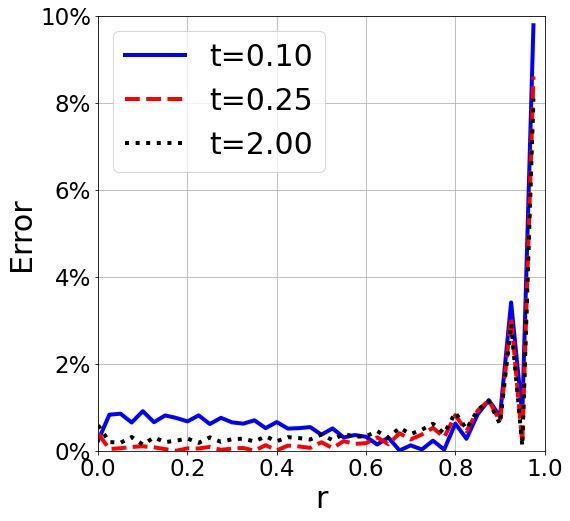}
         \caption{Relative percentage error, $\Delta t = 10^{-3}$.}
         \label{fig:Poiseuille-1D_error_32}
     \end{subfigure}
     \hfill
     \begin{subfigure}[b]{0.32\textwidth}
         \centering
         \includegraphics[width=\textwidth]{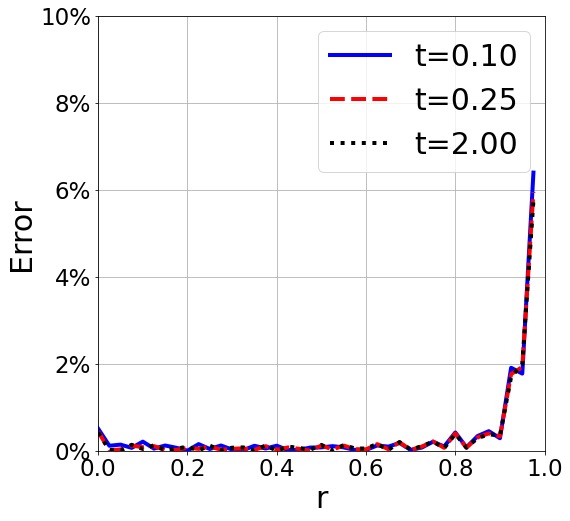}
         \caption{Relative percentage error, $\Delta t = 10^{-4}$.}
         \label{fig:Poiseuille-1D_error_32_dt=10^-4}
     \end{subfigure}

    \caption{Same as for Figure~\ref{fig:Poiseuille_1D_16} for $N=32$ radial modes.}
    \label{fig:Poiseuille_1D_32}
\end{figure}

Our simulation time-steps the prognostic equation in \eqref{eq:poiseuille} with the operator-splitting method described in \S\,\ref{sec:laplacian}. The constant pressure gradient is dealt with the forward Euler method, i.e., $\Delta u = G \Delta t$. Figures \ref{fig:Poiseuille-1D_evolution_16} and \ref{fig:Poiseuille-1D_evolution_32} show the time evolution of a slice of the velocity field with $GRe = 1$ and $\Delta t = 10^{-3}$. We use $N=16$ and $32$ radial modes, respectively. The blue triangles are the simulated values at pseudospectral grid points $\{r_{0,i}\}$. Figures \ref{fig:Poiseuille-1D_error_16} and \ref{fig:Poiseuille-1D_error_32} show the relative percentage errors with respect to the analytical solution, $|u_{\textrm{simulated}} - u_{\textrm{analytic}}|/u_{\textrm{analytic}}$. The error is below 4\% except in the boundary layer, where the velocity field in the denominator is close to zero. Increasing the number of radial modes $N$ yields a smoother error plot and a diminished the error boundary layer. However, the overall percentage is not significantly reduced. This is due to the error introduced by the Euler method, which we verified by reducing the time step (see Figures~\ref{fig:Poiseuille-1D_error_16_dt=10^-4} and~\ref{fig:Poiseuille-1D_error_32_dt=10^-4}).

\begin{figure}[htbp]
    \centering
    \includegraphics[width=90 mm]{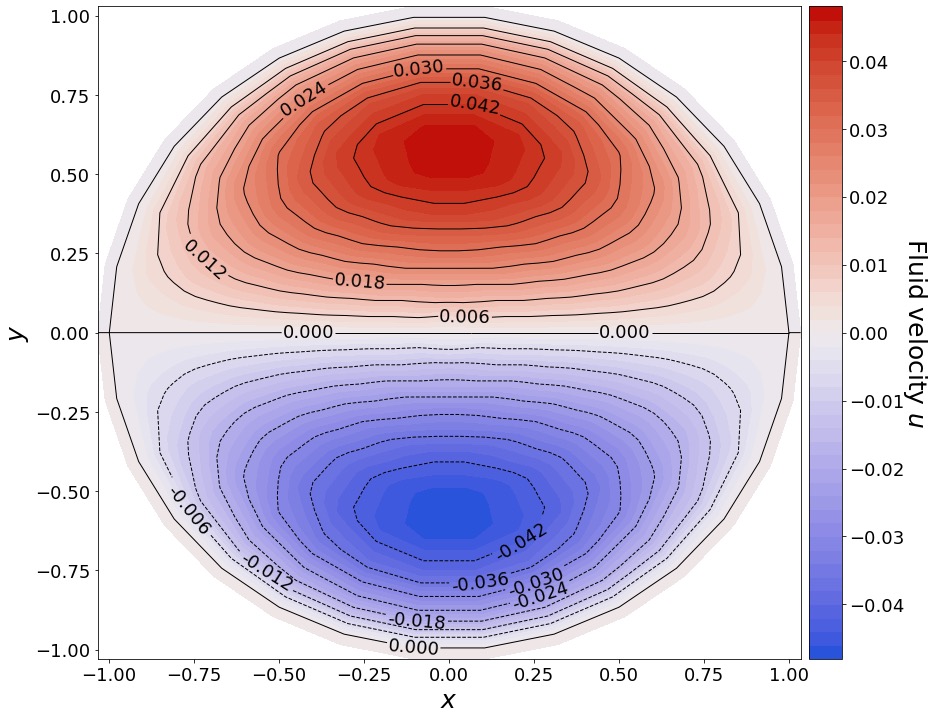}
    \caption{Steady solution to Eqn.~\eqref{eq:poiseuille} at $t=2$ for $GRe = y = r\sin{\theta}$.}
    \label{fig:Poiseuille-2-D_t=2}
\end{figure}

Figure~\ref{fig:Poiseuille-2-D_t=2} illustrates an example subjected to a nonconstant pressure gradient $G = y = r\sin{\theta}$ (varies along the vertical axis)  and $Re = 1$. The problem is now two-dimensional, and multiple angular modes are involved. Although no analytical solution is available for such a pressure gradient, the resulting steady velocity field satisfies the Dirichlet condition, and the solution respects the symmetry of the applied pressure gradient. With $N=15$ azimuthal and radial modes, and $\Delta t = 10^{-3}$, the simulation took a desktop Intel Core-i7 8700K processor 7.33 seconds to reach $t=2$, or 2000 iterations.



\subsection{Quantum vortices in 2-D Bose-Einstein condensate} \label{sec:BEC}
This example was studied by Grisouard~\cite{2009_GFD_Grisouard}, who laid out the basic features of our method. Numerous works have been published in the literature on the numerical methods of Bose-Einstein condensate (BEC)~\cite{Bao2006, Bao2009, Bao2012, Bao2014} and references therein. We recognize that the referenced works present more appropriate numerical techniques for addressing broader problems related to BECs. Nevertheless, the objective of our paper is to demonstrate the effectiveness of the method we have introduced, rather than pursuing superior solutions. We consider a BEC inside a circular infinite potential well. The dynamics of BECs are governed by the time-dependent Gross-Pitaevskii equation (GPE)~\cite{Gross:343403, pitaevskii1961vortex}, whose nondimensional form is given by
\begin{equation}
    \frac{\partial \psi}{\partial t} = \frac{\cplxi}{2}\nabla^2 \psi + \frac{\cplxi}{2\xi^2} (1-|\psi|^2)\psi,\quad\text{where}\quad \xi = \frac{\hbar}{R\sqrt{2mU_0 n_0}}
    \label{eq:GPENdim}
\end{equation}
is a dimensionless so-called ``healing length'' built from the system's parameters, with $\hbar$ the reduced Planck constant, $R$ the radius of the potential well that serves as our unit length, $m$ the particle mass, $U_0$ a constant potential representing repulsion between particles, and $n_0$ the normalization constant of the probability density function $|\psi|^2$. The details of nondimensionalization are further explained by Guo \& B\"{u}hler ~\cite{Guo2014}.
The external potential is set to be infinite (nullifying $\psi$) outside the unit radius ($R$ in dimensional units), such that the wave function $\psi$ satisfies the homogeneous Dirichlet condition.

Vortex circulation in a BEC is quantized and thus vortex strength cannot be continuously varied. We initialize a point vortex of circulation $\Gamma = 2\pi p$, $p\in \mathbb{Z}$, centered at $(r_0, \theta_0)$ by adding the wave function
\begin{equation}
    \psi_{\textrm{v}}^{\{r_0;\theta_0;p\}}(r,\theta) = \frac{r \eexp{\cplxi p\theta} - r_0 \eexp{\cplxi p \theta_0}}{\sqrt{\xi^2 + \left|r \eexp{\cplxi p\theta} - r_0 \eexp{\cplxi p \theta_0}\right|^2}}
\end{equation}
to an initial background wave function $\psi_{bg}$ that is meant to approximate the steady solution to the GPE, such that
\begin{equation}
    \psi(t=0) = \psi_{\textrm{bg}}\psi_{\textrm{v}}^{\{r_0;\theta_0;p\}}, \quad \textrm{where} \quad \psi_{\textrm{bg}} (r,\theta) = \tanh\left(\frac{1-r}{\sqrt{2}\xi}\right).
    \label{eq:background}
\end{equation}
The form, $\psi_{bg}$ takes, derives from the steady solution to the GPE in the case of an infinite, straight wall located at $x=0$, namely, $\psi_\infty = \tanh(x/2\xi)$ \cite{Mason2006}.
That is, a smooth transition between $\psi_\infty=0$ at the wall and $\psi_\infty \approx 1$ a few units of $\xi$ away from it.
In our circular domain, this solution is only approximate: having $\psi(t=0)=\psi_{bg}$ as our sole initial condition results in the creation of waves, oscillating about the exact and steady solution.
Here, we use $\xi=0.1$ and therefore, $\psi_{bg}$ is a good approximation of the steady solution to the GPE, i.e., radiation of transient waves is weak.

BEC dynamics share many similarities with classical shallow water systems, as revealed by the Madelung transformation \cite{Madelung1927}. Thus, vortex motion can be analyzed classically. For example, the time for a point vortex to loop around a unit disk can be calculated by using the method of images. As explained by Mason et al.~\cite{Mason2006}, due to the boundary layer effect caused by the infinite potential well, the actual free-slip boundary (virtual wall) for the method of images to be applied is a distance $\sqrt{2}\xi$ away from the potential wall.
The loop time is then given by~\cite{KUNDU2016195}
\begin{equation} \label{eq:theoretical_looptime}
    \tau_{\textrm{loop}} = \frac{4\pi^2 [(1- \sqrt{2}\xi)^2 - r_0^2]}{\Gamma}.
\end{equation}

\begin{figure}
    \centering
    \includegraphics[width = 90 mm]{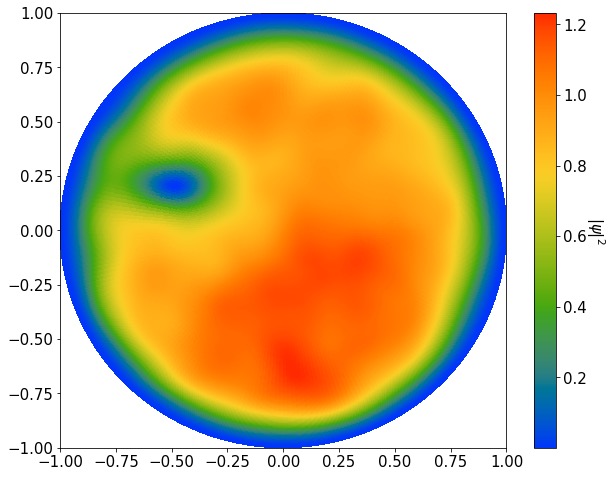}
    \caption{Probability density $|\psi|^2$ for a vortex with $r_0 = 0.6$ at $t=1$. The vortex is at the centre of the patch of lower values of $|\psi|^2$ in the top left quadrant.}
    \label{fig:BEC_waveform}
\end{figure}

We initialize a simulation with an initial wave function described by Eqn.~\eqref{eq:background}.
We set $\xi = 0.1$ and the size of the time step is $\Delta t = 5 \times 10^{-5}$. We sample the wave function with $256$ angular and radial points, so there are $N=128$ radial modes involved in the simulation, as discussed in \S\,\ref{sec:interpolation} and~\ref{sec:complexity}.
The nondimensional equation~\eqref{eq:GPENdim} is treated with the operator splitting method mentioned in \S\,\ref{sec:laplacian}.
For the nonlinear part, we have
\begin{equation}
    \frac{\partial \psi}{\partial t} = \frac{\cplxi}{2\xi^2}\left(1-|\psi|^2\right)\psi, 
\end{equation}
which can be solved analytically because this equation leaves $|\psi|$ unchanged. Indeed, the variations of $\psi$ go as $\cplxi K \psi$, where $K$ is real, meaning that only the phase of $\psi$ changes. Thus, taking $\psi_{\textrm{int}}$ as the initial condition, we can integrate over one time step as
\begin{equation}
    \psi(t+\Delta t) = \psi_{\textrm{int}} \exp \left[ \frac{\cplxi \left(1-|\psi|^2\right)}{2\xi^2} \Delta t \right],
\end{equation}
where we recall that $\psi_{\textrm{int}}$ is computed by Eqn.~\eqref{eq:time_scheme}. 
Figure~\ref{fig:BEC_waveform} shows $|\psi(t=1)|^2$ in the presence of a point vortex with $r_0 = 0.6$.

\begin{figure}
\centering
     \begin{subfigure}[b]{0.31\textwidth}
         \centering
         \includegraphics[width=\textwidth]{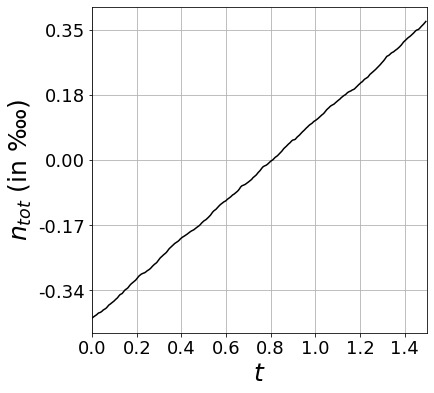}
         \caption{Total mass}
         \label{fig:BEC_n0}
     \end{subfigure}
     \begin{subfigure}[b]{0.31\textwidth}
         \centering
         \includegraphics[width=\textwidth]{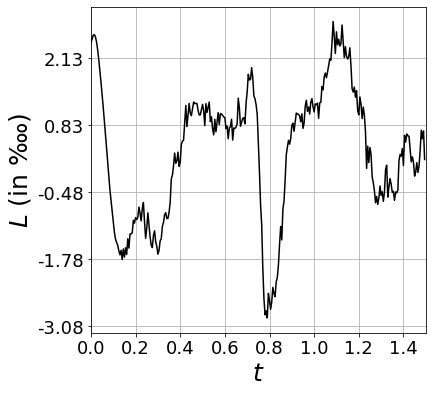}
         \caption{Angular momentum}
         \label{fig:BEC_angmom}
     \end{subfigure}
    \begin{subfigure}[b]{0.31\textwidth}
         \centering
         \includegraphics[width=\textwidth]{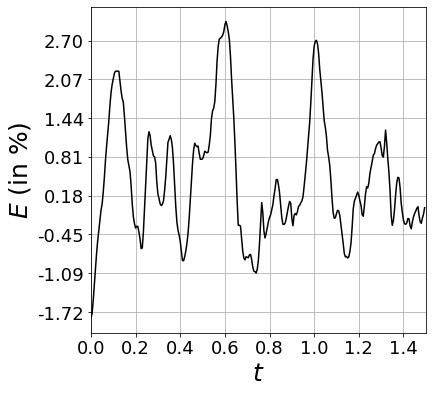}
         \caption{Energy}
         \label{fig:BEC_Energy}
     \end{subfigure}
    \caption{Evolution of the invariants for a BEC vortex initially centered at $r_0=0.6$.}
    \label{fig:BEC_conserve}
\end{figure}

Figure \ref{fig:BEC_conserve} shows the invariants of the vortex with $r_0=0.6$ over the time period $\Delta t=1.5$. The total mass $n_{\textrm{tot}}$, angular momentum $L$, and energy $E$ of the BEC are, respectively, \cite{2009_GFD_Grisouard, Guo2014}
\begin{subequations}
\begin{equation}
    n_{\textrm{tot}} = \int |\psi|^2 \di \textbf{a};
\end{equation}
\begin{equation}
    L = -\cplxi \int \psi ^* \frac{\partial \psi}{\partial \theta} \di \textbf{a},\quad \text{and}\quad
\end{equation}
\begin{equation} \label{eq:energy}
    E = \frac{1}{2} \int \left[|\nabla \psi|^2 + \frac{1}{2 \xi^2}\left(|\psi|^2 +1 \right)^2\right] \di \textbf{a}.
\end{equation}
\end{subequations}
The mass is constant within 0.4\textpertenthousand{} (Figure~\ref{fig:BEC_n0}), the angular momentum fluctuates within 0.5\textpertenthousand{} (Figure~\ref{fig:BEC_angmom}), while the energy fluctuates by about 2.7\% (Figure~\ref{fig:BEC_Energy}). The energy expression (Eqn.~\ref{eq:energy}) includes the most derivatives, both azimuthal and radial. The former derivative is calculated spectrally using FFTs, and the latter via a second-order finite-difference scheme. The latter suggests that the fluctuations might be due to the lack of precision of our postprocessing methods, rather than to that of the spectral method itself.

\begin{figure}
    \centering
    \includegraphics[width = \textwidth]{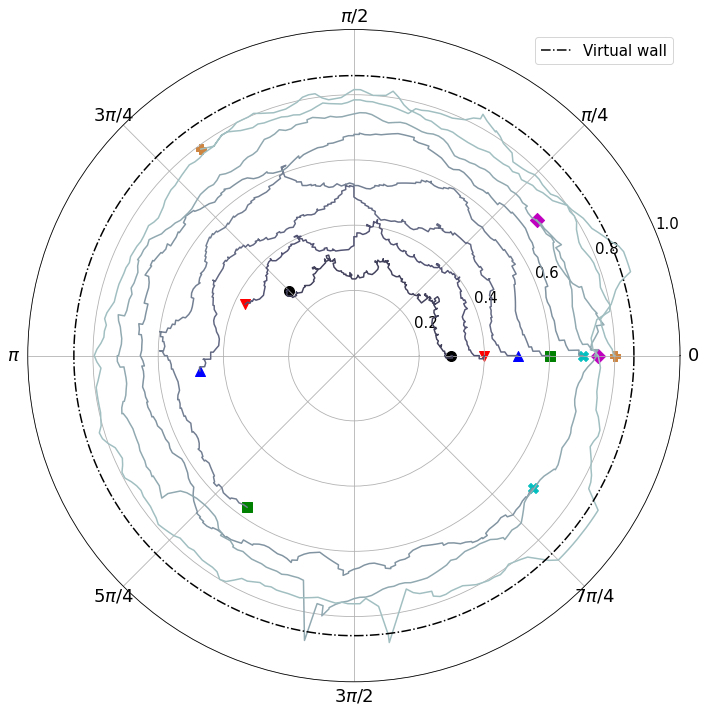}
    \caption{Traces of vortices' centers over a period of 1.5 unit time.}
    \label{fig:BEC_vortex_trace}
\end{figure}

We then run a suite of single-vortex simulations, each with a different initial radius for the vortex center: $r_0 = 0.3$, $0.4$, $0.5$, $0.6$, $0.7$, $0.75$, $0.8$ and $0.9$. All vortices start from $\theta_0 = 0$. Figure~\ref{fig:BEC_vortex_trace} shows the trajectories of the vortex centers over a period of time of $ \Delta t = 1.5$. The marker pairs show the start and end positions of the vortex centers. 
As mentioned earlier, the vortex centre moves parallel to the circular wall, describing circles around the domain center.
However, those circles are not perfect due to interactions with BEC waves.

\begin{figure}
    \centering
        \includegraphics[width = 90 mm]{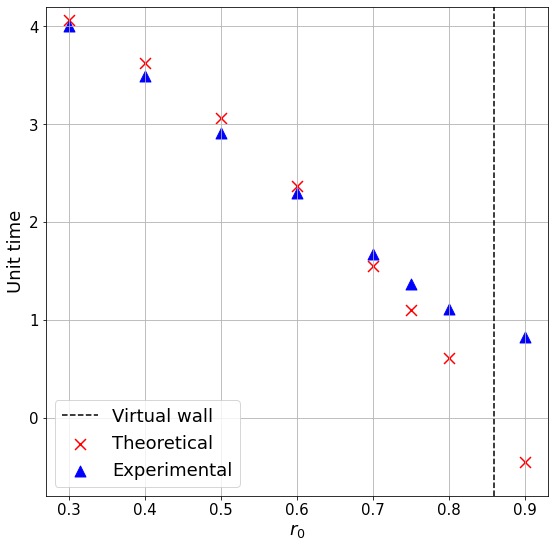}
    \caption{Theoretical loop time $\tau_0$ v.s. experimental values.}
    \label{fig:BEC_loop_time}
\end{figure}

Figure~\ref{fig:BEC_loop_time} shows measurements of the time $\tau_0$ it takes for a vortex center to describe a full circle around the center of the domain, versus the theoretical values predicted by the image method (Eqn.~\ref{eq:theoretical_looptime}). As expected, the experimental results deviate from the theoretical values when the vortex center is close to or inside the virtual wall at $r=1-\xi \sqrt{2}$. 

By comparing the looping times and computing the invariants, we validated our method on a nonlinear, pseudo-spectral simulation. This example also demonstrates the method at its full potential. Each simulation requires 30,000 iterations, which takes around 40 minutes to run on the same system as in \S\,\ref{sec:Poiseuille}. The method solves a PDE involving Laplacian on a complex field with $128\times 128$ angular and radial modes below 0.1~s.

\section{Conclusions and perspectives} \label{sec:conclusion}
This paper introduced a novel spectral method suited for Cauchy problems featuring Laplacians in polar coordinates. We reformulated the discrete Hankel transforms as pseudospectral methods, and we introduced DHTs' quadrature weights, cardinal functions, and pseudospectral grids. We discussed the choice of sampling grids for both FFTs and DHTs, and analyzed the error bounds of the numerical interpolation and DHTs. We demonstrated a systematic way to use the method to compute Laplacians and apply spectral time schemes to time-dependent partial differential equations. We then showed a range of time-dependent examples including linear and nonlinear PDEs, fields with complex numbers, and the 2-D wave equation involving the second order time derivative. We verified the theoretical error bound on functions with different boundary conditions. The method was validated by comparing the simulation results with analytical solutions and by checking energy, mass, and angular momentum conservations. Although all our equations and examples are on unit disks throughout this paper for concision, the method can be rescaled to arbitrary radii \cite{Baddour2019, 2004_JOSAA_Guizar-SicairosG}. We acknowledge that the literature has extensively studied the discretization of the Laplacian operator and Poisson-type equations in polar coordinates, employing finite difference methods, finite element methods, or alternative spectral methods with orthogonal polynomials. Many of these methods exhibit faster convergence rates, resulting in higher accuracy when computing Laplacians compared to the technique we present in this paper. Nevertheless, our method could potentially offer greater time marching efficiency due to the mathematically exact time marching scheme that utilizes eigenfunctions of the Laplacian (see \S~\ref{sec:laplacian}). However, a more detailed comparison is required to address this claim, and this also depends on the nonlinear parts.

A way to greatly speed up the method is to implement the program for parallel computing. After the FFT step, for which parallel libraries are widely available, each angular mode is  independent of the others, and the following interpolations and DHTs can be distributed to different processes at the same time. Due to the nature of Bessel functions, the method converges poorly if the homogeneous Dirichlet condition is not satisfied. However, there are many known ways to homogenize boundary conditions~\cite{BISWAS2019721, Friedman1997, Achdou1998}, which may be combined with our method for more general problems. We are currently considering a new transform based on Dini series, a variation of the Fourier-Bessel series which works on Neumann and Robin boundary conditions~\cite[\S\,18.11]{watson1995treatise}. A combination of these transforms has the potential to compute Laplacians under arbitrary boundary conditions. Holman and Kunyansky suggest a reduced polar sampling grid~\cite{Holman2015}, which becomes more sparse towards the domain center. Our current sampling grid has redundancy in the inner part at high angular modes, and adopting their strategy may improve efficiency.

\section*{Acknowledgment}
N.G. thanks the 2009 WHOI Geophysical Fluid Dynamics program on nonlinear waves and especially Oliver Bühler for fruitful scientific discussions.

\bibliography{novel_spectral.bib}

\appendix
\section{Proof of \texorpdfstring{$S_q^N(r) = f_q^N(r)$}{TEXT}} \label{sec:proof}

We prove the equality in Eqn.~\eqref{eq:equality}. By substituting the Fourier-Bessel expansion Eqn.~\eqref{eq:basisfunction} into the forward transform formula Eqn.~\eqref{eq:forward}, we have
\begin{equation}
    F_{q,j} = \frac{2}{k_{q,N+1}^2} \sum_{i=1}^{N}\frac{J_q\left(k_{q,j} r_{q,i}\right)}{J_{q+1}^2\left(k_{q,i}\right)}\left[ \sum_{m=1}^{\infty}a_{q,m} J_q\left(k_{q,m} r_{q,i}\right)\right].
\end{equation}
We rearrange this expression into
\begin{equation}
     F_{q,j} = \frac{2}{k_{q,N+1}^2}  \sum_{i=1}^N \sum_{m=1}^\infty \frac{a_{q,m}}{J_{q+1}^2\left(k_{q,i}\right)} J_q\left(k_{q,j} r_{q,i}\right)J_q\left(k_{q,m} r_{q,i}\right) .
\end{equation}
Summing the above equation over the pseudospectral grid points $\{r_{q,i}\}$, from the orthogonality relation \eqref{eq:orthogonality}, we find the expression we were looking for, namely,
\begin{equation}
    F_{q,j} =\frac{1}{2} \sum_{m=1}^\infty a_{q,m} J_{q+1}^2 \left(k_{q,m}\right) \delta_{jm} =  \frac{J_{q+1}^2(k_{q,j})}{2}a_{q,j}.
\end{equation}
Therefore, applying a DHT with $N$ radial modes is equivalent to truncating a Fourier-Bessel series at the $N^{\textrm{th}}$ term.

\end{document}